\documentclass[prl,twocolumn,showpacs,preprintnumbers,amsmath,amssymb]{revtex4}
\usepackage{bm}
\usepackage{amsmath}
\usepackage{dcolumn}
\usepackage[dvips]{graphicx}
\begin{document}
\bibliographystyle{apsrev}

\title{Radiative corrections and  parity nonconservation in heavy atoms}

\author{A. I. Milstein}
\email[Email:]{A.I.Milstein@inp.nsk.su} \affiliation{Budker
Institute of Nuclear Physics, 630090 Novosibirsk, Russia}
\author{O. P. Sushkov}
\email[Email:]{sushkov@phys.unsw.edu.au} \affiliation{School of
Physics, University of New South Wales, Sydney 2052, Australia}
\author{I. S. Terekhov}
\email[Email:]{I.S.Terekhov@inp.nsk.su} \affiliation{ Novosibirsk
University, 630090 Novosibirsk, Russia}

\date{\today}
\begin{abstract}
The self-energy and the  vertex radiative corrections to the
effect of parity nonconservation in heavy atoms are calculated
analytically in orders $Z\alpha^2$ and
$Z^2\alpha^3\ln(\lambda_C/r_0)$, where $\lambda_C$ and $r_0$ being
the Compton wavelength and the nuclear radius, respectively. The
value of the radiative correction is $-0.85\%$ for Cs and
$-1.41\%$ for Tl. Using these results we have performed analysis
of the experimental data on atomic parity nonconservation. The
obtained values of the nuclear weak charge,
$Q_W=-72.90(28)_{exp}(35)_{theor}$ for Cs, and
$Q_W=-116.7(1.2)_{exp}(3.4)_{theor}$ for Tl, agree with
predictions of the standard model. As an application of our
approach we have also calculated analytically dependence of the
Lamb shift on the finite nuclear size.

\end{abstract}
 \maketitle

Atomic parity nonconservation (PNC) has now been measured in
bismuth \cite{Bi}, lead \cite{Pb}, thallium \cite{Tl}, and cesium
\cite{Cs}. Analysis of the data provides an important test of the
standard electroweak model and imposes constraints on new physics
beyond the model, see Ref. \cite{RPP}. The analysis is based on
the atomic many-body calculations for Tl, Pb, and Bi \cite{Dzuba1}
and for Cs \cite{Dzuba2,Blundell} (see also more recent Refs.
\cite{Kozlov,DFG}). Both the experimental and the theoretical
accuracy is best for Cs. Therefore,  this atom provides the most
important information on the standard model in the low-energy
sector. The analysis performed in Ref. \cite{Cs} has indicated a
deviation of the measured weak charge value from that predicted by
the standard model by 2.5 standard deviations $\sigma$.

In the many-body calculations  \cite{Dzuba1,Dzuba2,Blundell} the
Coulomb interaction between electrons was taken into account,
while the magnetic interaction was neglected. The contribution of
the magnetic (Breit) electron-electron interaction was calculated
in the papers \cite{Der,DHJS}. It proved to be much larger than a
naive estimate, and it  shifted the theoretical prediction for PNC
in Cs.

Radiative correction to the nuclear weak charge due to
renormalization from the scale of the W-boson mass down to zero
momentum has been calculated long time ago, see Refs.
\cite{Mar1,Mar2}. This correction is always accounted in the
analysis of data. However, another important class of radiative
corrections was lost in the analysis of atomic PNC. This fact has
been pointed out in the recent work \cite{Sushkov} that
demonstrated that there are corrections $\sim Z\alpha^2$ caused by
the strong electric field of the nucleus. Here Z is the nuclear
charge and $\alpha$ is the fine structure constant. The simplest
correction of this type is due to the Uehling potential. It has
been calculated numerically in Ref. \cite{W} and analytically in
our paper \cite{MiSu}. In the paper \cite{MiSu} we have also
analyzed the general structure of the strong field radiative
corrections. It has been shown that except of the usual
perturbative parameter $Z\alpha$ there is an additional parameter
$\ln(\lambda_C/r_0)$ where $\lambda_C$ is the electron Compton
wavelength and $r_0$ is the nuclear radius.

In this letter we present the results of calculations of the
strong field radiative corrections to the atomic PNC effect in
orders $Z\alpha^2$ and $Z^2\alpha^3\ln(\lambda_C/r_0)$. Details of
calculations will be presented elsewhere \cite{Mil}. Using our
results we reanalize the experimental data. Agreement with the
standard model is excellent. As an application of our approach we
have also calculated the dependence of the Lamb shift on the
finite nuclear size. Agreement of our analytical formula with
results of previous computations \cite{CJS93,JS} is perfect.

\begin{figure}[h]
\centering
\includegraphics[height=90pt,keepaspectratio=true]{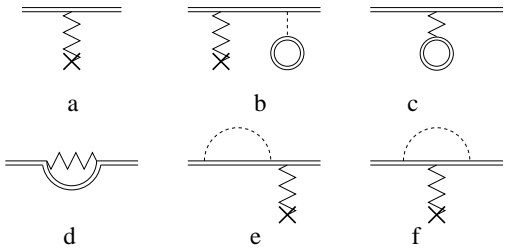}
\caption{\it (a) Leading contribution to the PNC matrix element.
(b-f) Radiative corrections. The double line is the exact electron
Green's function in the Coulomb field of the nucleus, the cross
denotes the nucleus, the zigzag and the dashed lines denote
Z-boson and photon, respectively.} \label{Fig1}
\end{figure}

The strong relativistic enhancement makes PNC radiative
corrections different from previously considered radiative
corrections to the hyperfine structure. The relativistic
enhancement factor is proportional to $R \sim (\lambda_C/(Z\alpha
r_0)^{2(1-\gamma)}$, where $\gamma=\sqrt{1-(Z\alpha)^2}$. The
factor is $R \approx$3 for Cs and $R \approx 9$ for Tl, Pb, and Bi
\cite{Kh}. The logarithmic enhancement of radiative corrections
mentioned above is closely related to the existence of the factor
$R$. The Feynman diagram for the leading contribution  to the PNC
matrix element between $p_{1/2}$ and $s_{1/2}$ states as well as
diagrams for radiative corrections are shown in Fig.\ref{Fig1}(a)
and Fig.\ref{Fig1}(b-f), respectively.
\noindent The diagram Fig.\ref{Fig1}(b) corresponds to a
modification of the electron wave function because of the vacuum
polarization. This correction calculated analytically in Ref.
\cite{MiSu} reads
\begin{equation} \label{db}
\delta_{b}=\alpha\left(\frac{1}{4}Z\alpha+
\frac{2(Z\alpha)^2}{3\pi\gamma}
\left[\ln^2(b\lambda_C/r_0)+f\right]\right),
\end{equation}
where $b=\exp(1/(2\gamma)-C-5/6)$, $C\approx 0.577$ is the Euler
constant, and $f\sim 1$ is some smooth function of $Z\alpha$
independent of $r_0$. Hereafter we denote by $\delta$ the relative
value of the correction. So, Eq. (\ref{db}) represents the ratio
of diagrams Fig.\ref{Fig1}(b) and Fig.\ref{Fig1}(a).

The renormalization of the nuclear weak charge $Q_W$ from the
scale of the W-boson mass down  to $q=0$ was performed in Refs.
\cite{Mar1,Mar2}. However, as it has been pointed out in Ref.
\cite{MiSu}, atomic experiments correspond to $q\sim 1/r_0\sim
30$MeV. The correction due to renormalization from $q=0$ to
$q=1/r_0$ is described by diagrams Fig.\ref{Fig1}(c) and
Fig.\ref{Fig1}(d). It has the form \cite{MiSu}
\begin{equation}
\label{cd} \delta_{cd}=\frac{4\alpha Z}{3\pi Q_W}
(1-4\sin^2\theta_W)\ln(\lambda_C/r_0)\sim -0.1\% \, ,
\end{equation}
where $\theta_W $ is the Weinberg angle, $\sin^2\theta_W\approx
0.2230$, see Ref. \cite{RPP}.

Diagrams Fig.\ref{Fig1}(e) and Fig.\ref{Fig1}(f) correspond to the
contributions of the electron self-energy operator and the vertex
operator, respectively. Each of these diagrams is not invariant
with respect to the gauge transformation of the electromagnetic
field while their sum  is gauge invariant. It has been
demonstrated in Ref. \cite{MiSu} that the correction $\delta_e
+\delta_f$ is of the form
\begin{equation}
\label{AB}
 \delta_{ef}=\delta_e +\delta_f=A\ln(b\lambda_C/r_0)+B \quad ,
\end{equation}
  where $A$ and $B$ are some functions of
$Z\alpha$, and the constant $b$ is defined after Eq.(\ref{db})  In
the present paper we derive the functions $A$ and $B$ analytically
in the leading approximation in the parameter $Z\alpha$, when
$A\propto \alpha(Z\alpha)^2$ and $B\propto \alpha(Z\alpha)$.

\begin{figure}[h]
\centering
\includegraphics[height=35pt,keepaspectratio=true]{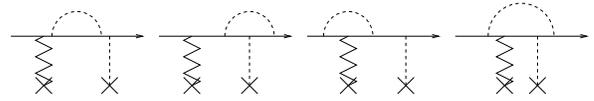} \caption{\it
{$\alpha(Z\alpha)$ self-energy and vertex radiative corrections.
The zigzag line denotes Z-boson, and the dashed line denotes
photon.}} \label{Fig2}
\end{figure}

The simplest part of the work is calculation of the
$\alpha(Z\alpha)$ contribution to $B$. It is convenient to use the
Fried-Yennie gauge \cite{FY} together with the effective operators
approach \cite{Eides} where the corrections under discussion
coincide with that  for the forward scattering amplitude, see
Fig.\ref{Fig2}.
The result of our calculation for the non-logarithmic term in
Eq.(\ref{AB}) reads
\begin{equation}\label{nlfinal}
B=-\alpha(Z\alpha)\left(\frac{7}{12}+2\ln 2\right).
\end{equation}

\begin{figure}[h]
\centering \vspace{10pt}
\includegraphics[height=35pt,keepaspectratio=true]{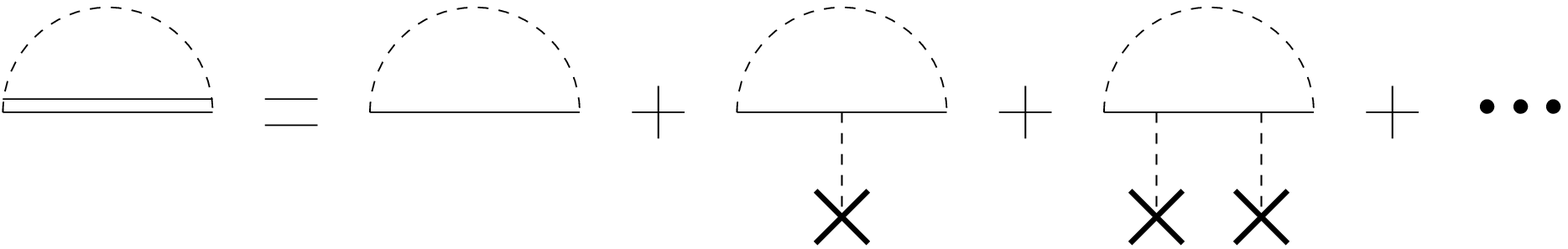} \caption{\it
{The electron self energy expanded in powers of the Coulomb field.
The solid line is the free electron Green's function, the cross
denotes the nucleus, and the dashed line denotes the photon. }}
\label{Fig3}
\end{figure}

For calculation of the function $A$ (the logarithmic part in Eq.
(\ref{AB})) we have used the Feynman gauge. There are two
contributions to $A$, the self-energy contribution, $A_{SE}$,
given by the diagram Fig.\ref{Fig1}(e), and the vertex
contribution $A_V$ given by Fig.\ref{Fig1}(f). It is convenient to
represent the self-energy operator as a series in powers of the
Coulomb field of the nucleus,
$\bm\Sigma=\bm\Sigma_0+\bm\Sigma_1+\bm\Sigma_2+...$, see
Fig.\ref{Fig3}, and perform  calculations in the momentum space.
\noindent
  Distances $r_0 \ll r \ll \lambda_C$ where the logarithmic terms
  come from correspond to momenta $m \ll p \ll 1/r_0$, where $m$
  is the  electron mass. Therefore, the mass and the electron
  binding energy can be neglected in electron propagators.
It is necessary to make the ultraviolet  regularization of the
operators $\bm\Sigma_0$ and $\bm\Sigma_1$. Each of the operators
depends on the parameter of the regularization. However,  the
contribution of $\bm\Sigma_{01}=\bm\Sigma_0+\bm\Sigma_1$ to
$A_{SE}$ is independent of the regularization parameter because of
the Ward identity. The operator $\bm\Sigma_2$ does not require any
regularization. Our results for contributions of $\bm\Sigma_{01}$
and $\bm\Sigma_2$ to $A_{SE}$ read
\begin{eqnarray}\label{munu}
A_{01}&=&-\frac{\alpha(Z\alpha)}{\pi}\quad ,\nonumber \\
A_{2}&=&-\frac{\alpha(Z\alpha)}{6\pi}(\pi^2-9)\quad ,\nonumber\\
A_{SE}&=&A_{01}+A_{2}=-\frac{\alpha(Z\alpha)^2}{6\pi}(\pi^2-3)\quad.
\end{eqnarray}

\begin{figure}[h]
\centering
\includegraphics[height=180pt,keepaspectratio=true]{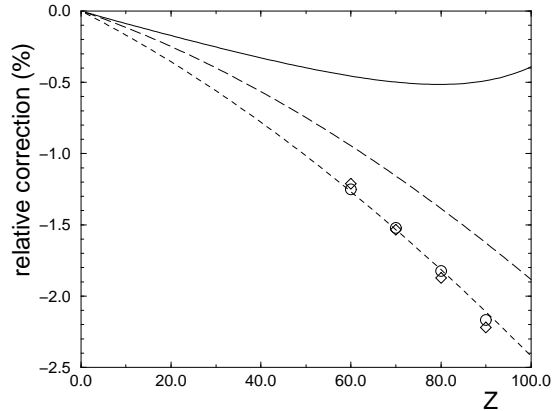}
\vspace{-10pt} \hspace{0pt} \caption{\it { Relative radiative
corrections (\%) for the PNC and for the finite-nuclear-size
effects versus the nuclear charge $Z$. The long-dashed line shows
the correction $\delta_{ef}$ given by Eq.(\ref{finaltot})
(Fig.\ref{Fig1}(e,f)). The solid line shows the total radiative
correction to the PNC effect that includes both $\delta_{ef}$ and
$\delta_{b}$ (Eq.(\ref{db}), Fig.\ref{Fig1}(b)). The dashed line
shows the correction $\Delta_s$ given by Eq.(\ref{finaFNSS}).
Results of computations of $\Delta_s$ for $1s$ and $2s$ states
\cite{CJS93,JS} are shown by circles and diamonds, respectively.
 }}
\label{Fig4}
\end{figure}

Next is the logarithmic contribution of the vertex operator
described by Fig.\ref{Fig1}(f). The coordinate representation is
the most convenient for this part of the problem, and we use this
representation. We have also used an integral representation of
the Green function derived in Ref.\cite{MS82} (see also Ref.
\cite{LeeM}). Keeping in mind the contact nature of the PNC
interaction and the logarithmic accuracy of the  calculation, one
can demonstrate that only the angular momentum $j\, =\, 1/2$ is
significant in the partial expansion of the electron Green's
function in the external Coulomb field. A part of the vertex
independent of $Z\alpha$ requires an ultraviolet regularization
and hence is dependent of the regularization parameter. However,
due to the Ward identity, the contribution  corresponding to this
part cancels out exactly the renormalization of the wave function,
see e.g. Ref. \cite{Pach96}. Another part is dependent on
$Z\alpha$ and gives the following contribution to the logarithmic
term in Eq. (\ref{AB})
\begin{eqnarray}\label{fanswer}
A_V=-\frac{\alpha(Z\alpha)^2}{\pi}\left(\frac{17}{4}-\frac{\pi^2}{3}\right)\quad.
\end{eqnarray}
Together with Eq. (\ref{munu}) this gives the final result for $A$
\begin{eqnarray}\label{finallog}
A=-\frac{\alpha(Z\alpha)^2}{\pi}\left(\frac{15}{4}-\frac{\pi^2}{6}\right)\quad.
\end{eqnarray}
Thus, according to Eqs. (\ref{AB}), (\ref{nlfinal}), and
(\ref{finallog}) the total relative correction to the PNC matrix
element due to the self-energy and the vertex operators reads
\begin{eqnarray}\label{finaltot}
\delta_{ef}&=&-\alpha\left[(Z\alpha)\left(\frac{7}{12}+2\ln2\right)\right.\nonumber\\
&+&\left.\frac{(Z\alpha)^2}{\pi}\left(\frac{15}{4}-\frac{\pi^2}{6}\right)
\ln(b\lambda_C/r_0) \right]\quad.
\end{eqnarray}
This correction is plotted in Fig. \ref{Fig4} versus the nuclear
charge $Z$ (the long-dashed line). The solid line in the same
figure shows the total radiative correction to the PNC effect that
includes both $\delta_{ef}$ (Eq. (\ref{finaltot}) and $\delta_{b}$
(Eq. (\ref{db})).

The radiative shift of the atomic energy levels (Lamb shift)
depends on the finite nuclear size. This correction has a very
similar structure to that of the PNC radiative correction since
the effective sizes of the perturbation sources in both cases are
much smaller than $\lambda_C$. The self-energy and the vertex
corrections to the finite-nuclear-size effect (SEVFNS) for
$s_{1/2}$-state have been calculated earlier analytically in order
$\alpha(Z\alpha)$ in Refs.\cite{Pach93,PG}. The corrections for
$1s_{1/2}$-, $2s_{1/2}$-, and $2p_{1/2}$-states have been
calculated numerically exactly in $Z\alpha$ in
Refs.\cite{CJS93,Blun92,LPSY}. The structure of higher in
$Z\alpha$ corrections and their logarithmic dependence on the
nuclear size has not been understood.
 We have applied our approach to the SEVFNS problem
 and found the following expression for the $s_{1/2}$-state
 relative correction
\begin{eqnarray}\label{finaFNSS}
\Delta_s&=&-\alpha\left[(Z\alpha)\left(\frac{23}{4}-4\ln2\right)\right.\nonumber\\
&+&\left.\frac{(Z\alpha)^2}{\pi}\left(\frac{15}{4}-\frac{\pi^2}{6}\right)
\ln(b\lambda_C/r_0) \right]\quad.
\end{eqnarray}
Linear in $Z\alpha$ term agrees with results of Refs.
\cite{Pach93,PG}. The logarithmic term coincides with that in Eq.
(\ref{finaltot}) for the PNC correction. Moreover, the logarithmic
term in the SEVFNS correction $\Delta_p$ for $p_{1/2}$ state is
also equal to that in Eqs.(\ref{finaltot}) and (\ref{finaFNSS}).
The reason for this equality is very simple. The logarithmic terms
come from small distances ($r << \lambda_C$) where the electron
mass can be neglected. When the mass is neglected the relative
matrix elements for the PNC radiative correction and for SEVFNS
are equal. The correction $\Delta_s$ given by Eq. (\ref{finaFNSS})
is shown in Fig. \ref{Fig4} by the dashed line. Results of the
computations \cite{CJS93,JS} for $1s$ and $2s$ states are shown by
circles and diamonds, respectively. The agreement is perfect.
\noindent
  It is interesting to find also the $\alpha (Z \alpha)$-term in
  $\Delta_p$. Using our result for the logarithmic term in
  $\Delta_p$ and results of numerical calculations of this
  correction for $2p_{1/2}$-state, Ref. \cite{CJS93,JS}, we find
  that the coefficient in $\alpha (Z \alpha)$-term is $-1.9$.

It has been recently suggested in Ref. \cite{K} that the following
"precise relation" $\delta_{ef}=(\Delta_s+\Delta_p)/2$ is valid.
Values of the corrections obtained in our work do not agree with
this relation. The "derivation" in Ref. \cite{K} is based on the
obviously wrong assumption that there is a gauge in which the
vertex contributions to $\delta_{ef}$, $\Delta_s$, and $\Delta_p$
vanish simultaneously. Although it is possible to vanish the
vertex correction to each of these quantities by chousing an
appropriate gauge, these are three different gauges. This is why
the above relation is wrong.

Now we can perform a consistent analysis of the experimental data
on atomic parity violation since all the contributions are known.
In our analysis for Cs we have included the theoretical value of
the PNC amplitude from Refs. \cite{Dzuba2,Blundell,Kozlov,DFG} as
well as $-0.61\%$ correction due to the Breit interaction
\cite{DHJS}, $-0.42\%$ radiative correction calculated in the
present work, $-0.2\%$ neutron skin correction \cite{Der1},
$-0.08\%$ correction due to the renormalization of $Q_W$ from the
atomic momentum transfer $q\sim 30$MeV down to $q=0$ \cite{MiSu},
and $+0.04\%$ contribution from the electron-electron weak
interaction \cite{MiSu}. Using these theoretical results we obtain
from the data \cite{Cs} the following value of the nuclear weak
charge $Q_W$ at zero momentum transfer
\begin{equation}
\label{QCs} Cs: \quad Q_W=-72.90 \pm (0.28)_{ex}\pm(0.35)_{theor}.
\end{equation}
This value agrees with prediction of the standard model,
$Q_W=-73.09\pm 0.03$, see Ref. \cite{RPP}. We have used the
neutron skin correction  in our analysis. However, in our opinion,
status of the correction is not quite clear because data on the
neutron distribution used in Ref. \cite{Der1} are not quite
consistent with the data on neutron distributions obtained from
proton scattering, see e. g. Ref. \cite{LIYAF}.

In the analysis for Tl we have included the theoretical value of
the PNC amplitude from Refs. \cite{Dzuba1}, as well as $-0.88\%$
correction due to the Breit interaction \cite{rescale}, $-0.51$\%
radiative correction calculated in the present work, $-0.2\%$
neutron skin correction, $-0.08\%$ correction due to the
renormalization of $Q_W$ from the atomic momentum transfer $q\sim
30$MeV down to $q=0$ \cite{MiSu}, and $+0.01\%$ contribution from
the electron-electron weak interaction \cite{MiSu}. Using these
theoretical results we obtain from the data \cite {Tl} the
following value of the nuclear weak charge $Q_W$ at zero momentum
transfer
\begin{equation}
\label{QTl} Tl: \quad Q_W=-116.7 \pm (1.2)_{ex}\pm(3.4)_{theor}.
\end{equation}
This  agrees with prediction of the standard model, $Q_W=-116.7\pm
0.1$, see Ref. \cite{RPP}.

Concluding, we have calculated analytically for the first time all
the strong electric field radiative corrections to the effect of
atomic parity violation. This calculation has allowed us to
perform a consistent analysis of the experimental data. Agreement
with the standard model is within $0.5\sigma$.

A.I.M  gratefully acknowledge School of Physics at the University
of New South Wales  for  warm hospitality and financial support
during a visit.\\


\begin{thebibliography}{99}
\bibitem{rescale} We found this value by rescaling the Breit
correction for Cs \cite{DHJS}, see discussion in Refs.
\cite{Sushkov,MiSu}.
\end{thebibliography}
\end{document}